UNIVERSITY OF CALIFORNIA
RIVERSIDE

An Analogy Based Method for Freight Forwarding Cost Estimation

A Thesis submitted in partial satisfaction
of the requirements for the degree of

Master of Business Administration

in

Management

by

Kevin Andrew Straight

June 2014

Thesis Committee:
      Dr. Mohsen El Hafsi, Chairperson
      Dr. Yunzeng Wang
      Dr. Elodie Goodman

The Thesis of Kevin Andrew Straight is approved:

_______________________________________________

_______________________________________________

_______________________________________________
                                    Committee Chairperson

University of California, Riverside

**Acknowledgements**

I would like to thank Seamus McGowan, Aidan Conway, and the rest of the management staff of Independent Express Cargo in Dublin Ireland for their kind help and patience while I was conducting the field portion of my research.

I would especially like to thank Danica Sheridan, MLIS, who has not only been a loving a supportive partner, but has also been invaluable as a reference librarian and copy editor throughout my program.

ABSTRACT OF THE THESIS

An Analogy Based Method for Freight Forwarding Cost Estimation

by

Kevin Andrew Straight

Master of Business Administration, Graduate Program in Management
University of California, Riverside, June 2014
Dr. Mohsen El Hafsi, Chairperson

The author explored estimation by analogy (EBA) as a means of estimating the cost of international freight consignment. A version of the k-Nearest Neighbors algorithm (k-NN) was tested by predicting job costs from a database of over 5000 actual jobs booked by an Irish freight forwarding firm over a seven year period. The effect of a computer intensive training process on overall accuracy of the method was found to be insignificant when the method was implemented with four or fewer neighbors. Overall, the accuracy of the analogy based method, while still significantly less accurate than manually working up estimates, might be worthwhile to implement in practice, depending labor costs in an adopting firm. A simulation model was used to compare manual versus analytical estimation methods. The point of indifference occurs when it takes a firm more than 1.5 worker hours to prepare a manual estimate (at current Irish labor costs). Suggestions are given for future experiments to improve the sampling policy of the method to improve accuracy and to improve overall scalability.

# CONTENTS

# LIST OF FIGURES

# LIST OF TABLES

# INTRODUCTION

International freight forwarding, the process of arranging the shipping, warehousing, loading, and documentation of cargo for third parties, plays an important role in many supply chains. Many freight forwarding jobs are sold through a competitive bidding process. The firm importing or exporting a consignment requests quotations from multiple freight forwarders who then bid on the consignment. The costs to move a consignment can vary widely based on the start and end points, the load size, and which subcontractors and modalities are used to transport the freight. Accurate information about costs and margins on particular routes can be an important source of competitive advantage for freight forwarders in terms of winning bids and earning the best possible profit margins on jobs. The ability to rapidly retrieve this information is also important as it allows the preparation of more bids in any given period ("Freight forwarder", n.d.).

One approach to estimating shipping costs for the preparation of freight forwarding bids is to refer to records of past jobs. These records, which are kept by all freight forwarders for invoicing and insurance purposes, contain a large amount of information about each job. The main challenge in estimating freight costs by analogy to previous jobs is to accurately classify historic jobs so that the most analogous jobs can be found. This would be a daunting prospect using manual methods because of the large number of job attributes and the large number of historic jobs. It is desirable to automate the process using computer pattern recognition. It is also desirable to add a "learning" element to the algorithm. That is, as more jobs are added to the database, the algorithm should be able to improve in accuracy over time.

*k*-Nearest Neighbors (*k*-NN) is a popular machine learning algorithm which is often used for problems of this type. *k*-NN returns the *k* records in the database which most closely resemble the entity in question. It lends itself to estimation problems because it is relatively simple to take a weighted average of the cost of the *k* previous jobs and use it as the estimate. Furthermore, it is also easy to modify the relative importance of the attributes by assigning different weights (Li, Xie, & Goh, 2009).

The *k*-NN algorithm belongs to a larger family of *lazy-learning* algorithms. Lazy learning algorithms have the advantage that they become more accurate as more data is added and that they require no assumptions or hypotheses about the functions or distributions of the data (Illa, Alonso, & Marré, 2004).

This project will describe an implementation of the *k*-NN algorithm to estimate freight costs and evaluate the performance of the method by testing it using a database of actual historical freight forwarding jobs.

## Importance of the Topic

Cost estimation is an important task in many industries. The ability to generate quality estimates in time for customer imposed deadlines is essential. An estimate that is too high may result in losing the job, whereas an estimate that is too low can force the firm to take a loss on the job. For large jobs or unusual jobs it can be cost effective for an experienced employee to solicit bids from subcontractors and calculate a detailed schedule of costs from the bottom up. In the case of repetitive jobs, a firm will often have negotiated rate agreements with its customers and suppliers so that the cost and revenue stay the same for long periods of time.  Between these two extremes, however, fall many jobs which are too small to warrant creation of a detailed manual estimate, yet too unique to be covered by a negotiated rate agreement.  Thus, it is highly desirable to have an automated system in place to generate estimates for these types of jobs.

While this particular project focuses on estimation in a freight forwarding context, the basic methodology could be applicable to cost estimation in other industries, such as construction or software development.

## PREVIOUS WORK

Nearest neighbor-type machine learning techniques are widely used for pattern recognition and data mining in fields as diverse as epidemiology, handwriting recognition, and credit card fraud detection. As early as the 1960's, authors such as Thomas Cover (1968) proposed their use in cost estimation and attempted to quantify the error bounds and risk function attendant in such a use.

In more recent years there seems to be increased interest in the use of $k$-NN techniques for cost estimation. Auer et a.l (2007) proposed a nearest neighbor model to estimate the cost of software projects. Their model considered the price of a software project as a function of the features it incorporated. They used a variation of the $k$-NN algorithm to estimate the cost of new projects by finding the $k$ most similar projects in a portfolio. The authors also devoted considerable attention to choosing optimum weights to each feature. However, they considered each feature of a project as a Boolean parameter (the project either had the feature or did not), whereas the attributes of freight forwarding jobs are scalars (time and load size) and vectors (collection and delivery location). Additionally, they did not fully address the role of time in project costs.

Li, Xie, and Goh (2009) also explored analogy based estimation for software features. They proposed a modified nearest neighbors approach in which the size of the data set was reduced by pruning before applying the $k$-NN search, resulting in much more efficient use of computer resources. Their paper is particularly interesting because it compares the fitness of estimates using several different similarity and solution functions and with four different variations of the k-NN algorithm. Like Auer, however, they make no attempt to correct for trends in project cost over time.

More recently Alkhatib et al. (2013) used a *k*-NN methodology to predict securities prices on the Jordanian stock exchange. Stocks, like freight costs, are functions of a complex, chaotic system that depends on both economic and human factors. However, the attributes they actually used to predict stock price (current price, opening price, closing price, high, and low) were all intrinsic to the stock price time series instead of being "real world" values like distance or load size. Additionally, stock prices are an essentially continuous time series whereas weeks or even years can pass between freight forwarding jobs to a given region.

One feature of analogy based methods is that weights can be assigned to different attributes to reflect their relative importance in the estimate. Different authors have suggested a number of approaches for how to assign these weights. Some authors (Shepperd & Schofield, 1997, p. 738) favor assigning weights manually based on the judgment of an expert. Others use a traditional grid search, in which different combinations of parameter values are tested at set intervals. Currently, the literature has tended to favor stochastic methods (Bergstra & Bengio, 2012).

Bergstra and Bengio argue that randomly guessing the parameter values tends to be preferable to either manually assigning the weights or a grid search method for high dimensional training problems. They imply that a high dimensional problem, in this context, is one with three or more parameters. Manually assigning weights, they argue, produces non-reproducible results because it relies too much on intuition and expert knowledge. Furthermore, experience and intuition are not always effective given that weights can vary greatly when the same method is used on different data sets. A grid search, on the other hand, tends to be an inefficient use of processor time. This is because the loss function of the algorithm tends to be much more sensitive to some parameters than others. Thus, the search spends many cycles testing combinations of parameters which have little effect on the overall error, while more important attributes are held constant.

Random search methods, they argue, generate more useful solutions in fewer cycles because more values of each parameter are tested. They also point out that random searches are particularly well suited to implementation on clusters of computers where some of the computers might fail during the process. Random searches also have the advantage that it is easy to add more trials or ignore the results of some trials. In their experiments, which consisted of finding optimum parameter weights for a classification algorithm based on various data sets, random searches returned better solutions than grid searches in the same amount of processing time in the majority of trials.

Another popular stochastic method relies on a genetic algorithm (GA) to set attribute weights. A starting value for each attribute weight is encoded as a "chromosome", typically made up of at least $2^{14}$ bits of information. The chromosomes are then randomly selected and paired with other chromosomes to produce new chromosomes. Additionally, a mutation effect is produced in the new population in such a way that each bit has a small chance of randomly mutating in each iteration. An elitist criterion ensures that the least fit combinations are eliminated from the population (Li,

Xie, & Goh, 2009, pp. 244-246).  While effective, GA is computationally intensive and was not considered feasible for this study, given available IT resources.

# MATERIALS AND METHODS

The general approach relied on a *nearest neighbors* algorithm. The pertinent attributes—date, collection coordinates, delivery coordinates, and load size—of each record in the database were mapped as a vector in 6 dimensional space. For each job estimated, a search function found the $k$ closest jobs in 6-space. The estimated cost for the job is the weighted average cost of the $k$ nearest jobs, normalized by load size, the weights having been determined in a prior training phase. A Euclidean distance formula is used to determine distance between the jobs. A slight complexity arose when dealing with the collection and delivery locations, which are coded as spherical, rather than Cartesian coordinates. The haversine formula (Equation 1), long used by navigators, was used to find the great circle distance between each set of collection or delivery coordinates (Sinnott, 1984).

$$d(p_1, p_2) = \arcsin\left(\sqrt{\sin^2\left(\frac{lat_2 - lat_1}{2}\right) + \cos(lat_1) \cdot \cos(lat_2) \cdot \sin^2(\frac{lng_2 - lng_1}{2})}\right) \qquad (1)$$

*where*:

$d(p_1, p_2)$ = the great circle distance between two points, $p_1$ and $p_2$

$lat_i$ = the latitude of point $i$

$lng_i$ = the longitude of point $i$

The attribute distance was then calculated using Equation 2. The constant weights in the equation represent the relative importance of different attributes, as determined in the training phase.

$$D = \sqrt{x_1 \cdot d(col_1, col_2)^2 + x_2 \cdot d(del_1, del_2)^2 + x_3 \cdot (t_2 - t_1)^2 + x_4 \cdot (l_2 - l_1)^2} \qquad (2)$$

*where*:

$D$ = the attribute distance between two consignments

$t_i$ = the date of consignment $i$, in days from a datum

$l_i$ = the load size of consignment $i$, as a fraction of a full truck load

$x_1, x_2, x_3, x_4$ = weights reflecting the relative importance of each attribute, as determined during a training phase

Once the list of the $k$-nearest jobs is found, a solution function (Equation 3) is used to estimate the normalized cost of the proposed job. This solution function returns the mean of the normalized costs of the $k$-nearest jobs, weighted by their distance from the job to be estimated. Normalized cost is the cost of the job divided by its load size. It has units of Euros/container. The estimated cost of the job to be estimated was found by multiplying the estimated normalized cost by the load size. An overview of this method is shown in pseudo code form in Figure 1.

$$S = \sum_{i=1}^{k} \left( \frac{D_{p,i}}{\sum_{j=1}^{k} D_{p,j}} \right) \left( \frac{cost_i}{l_i} \right) \qquad (3)$$

*where*:

$D$ = the attribute distance in between two consignments

$cost_i$ = the historical cost of job $i$

$l_i$ = the load size of job $i$, as a fraction of a full truckload

```
Let Jp = a job to be estimated

Let J[1 to n] = a table of historical jobs

KNearest[1 to k] = a table of the first k jobs

for i from k+1 to n

        find distance between Jp and J[i]

        if distance is less than any job in KNearest, replace the job in KNearest

next i

Cost = Weighted average by distance of cost of each KNearest
```

*Figure 1. k-*NN implementation for estimating job cost.

## Data Set

The data set to be used is a log of freight forwarding transactions from a third party logistics (3PL) firm based in Dublin, Ireland.  The logs cover the period June 22,

2006 to August 28, 2013.  These transactions are grouped into either "exports" or

"imports" for the sake of operational convenience, as the company's other functional

departments deal with each differently (e.g., for an export the company typically needs to

dispatch trucks to collect the freight whereas for an import they need to keep track of

when the ship is expected and book in delivery of the container).  The export logs also

include some intra-country consignments which were arranged by the freight forwarding

department.  The logs do not include the majority of the firm's small consignments

(approximately six pallets or less) between The Republic of Ireland and North Ireland or the United Kingdom, as the company has established relationships with pallet networks in these countries and they are arranged though a dedicated electronic data interchange (EDI) system.

The database contained records of 3,251 import consignment and 4,491 export consignments, for a total of 7,742. After canceled jobs, duplicate records, and incomplete records were removed, 6,745 records remained. Of these records, 5,035 had address information that were successfully coded as latitude and longitude coordinates.

## Materials

The data was initially received in a Microsoft Access database format. Accordingly, initial coding and cleaning of the data was performed on a laptop computer using a custom interface written in the VBA programming language over Microsoft Access, with calls to Google's *Geocoding* and *Maps* application programmer interfaces (APIs).

In the interests of speed and scalability, the training procedure and the actual experiments were implemented in the Python programming language on a specially constructed Linux based computer cluster composed of 16 processing cores spread across three commodity-class rack servers. The scripts made heavy use of the popular *SciPy* scientific computing and *NumPy* numeric analysis libraries[1].

Final data analysis was performed on a Windows laptop using *Microsoft Excel* and *IBM SPSS*.

---

[1] Both libraries are available for free download from http://www.scipy.org.

## Data Coding

The original database was created over a period of several years by multiple operators entering information into computer spreadsheets. There seems to have been little or no attempts at data validation or standardized formatting; different operators used different units and abbreviations and used different cells to record information. Luckily, at the start of this project, the data had already been converted to a relational database as part of a previous project, which included manually converting the three currency systems used (GBP, USD, and Euros) into Euros and formatting the records into a table structure, and converting many text strings to numbers (Straight, 2013). However, it was still necessary to code the load size and the collection and delivery latitudes and longitudes.

Data coding was one of the more labor intensive portions of the project, requiring over 400 man-hours over a period of approximately five months.

### Load Size

Load size is quoted in the original freight forwarding logs in a variety of units including pieces, cartons, standard pallets, euro-pallets, loading meters, kilograms, tons, truck-loads, and raw dimensions. Fundamentally, however, each of these units can be expressed as a fraction of the capacity of a standard shipping container. For purposes of this study, a standard shipping container is assumed to have a capacity of 26 standard pallets, 33 euro-pallets, 13.6 loading meters, or 24,000 kgs and these can be taken as approximate conversion factors.  These conversions were chosen because they are "rule of thumb" numbers that are actually used by the freight forwarding department of the company which supplied the data set (Conway, Aidan, 2013, July 29, personal communication).  While these simplifying assumptions resulted in some loss of precision, accuracy was maintained.   The goal was to compare the costs of similar loads.  For instance, the exact number of half pallets that fit in a container was relatively unimportant because consignments of a half pallet were unlikely to be nearest neighbors to full container consignments.  As long as ordinality was maintained and coding was applied consistently, the load size attribute would still have been relevant for calculating attribute distance.

Records were discarded if they could not reasonably be assigned a load size.  For instance, consignments recorded as "3 pieces", "one tractor", or "partial load" simply did not have enough information to code.

**Collection and Delivery Location**

The process of converting a place name or postal address to its corresponding latitude and longitude is known as *geocoding* ("Geocoding," 2014).  This project used Google's geocoding service, which offers an online applications programmer interface API that accepts a text string containing an address and returns the coordinates. Unfortunately, the majority of the address strings in the database were not accepted by Google's service because of misspellings, nonstandard abbreviations, missing information, or problems with character conversion between languages.  In most cases, the geocoding process required doing a web search for the website or social media page of the entity to find their mailing address, pasting the address into Google Maps to test it, replacing foreign characters with their nearest English equivalent[2], and finally passing the address string to the Maps API to be geocoded.

Records were flagged and discarded if their collection and delivery locations could not be resolved to within 25km.  In many cases, however, the coordinates were eventually resolved to the city block, or even individual building level.

## Data Segmentation

For all experiments, the data set was segmented into three subsets.  The first of

these, the test set, was composed of the most recent third of transactions and was retained

for the experiments.  The remaining two thirds of the data were randomly assigned to a

historical set and a training set.  While some authors, such as Li (2009), recommend an

equal split between the historical and training sets, this study follows the more common

practice of a 60%-40% split (Wilson & Keating, 2009, p. 449) (Figure 2).  The historical

set was used to predict the costs of the training set during the training phase.  The historical and training sets were then combined and used to predict costs of the test set during the experimental phase.

---

[2] For example, the word "Straße" ("street") appears in many German and Austrian addresses.  The Google
API would accept it properly after it was edited to "StraBe".

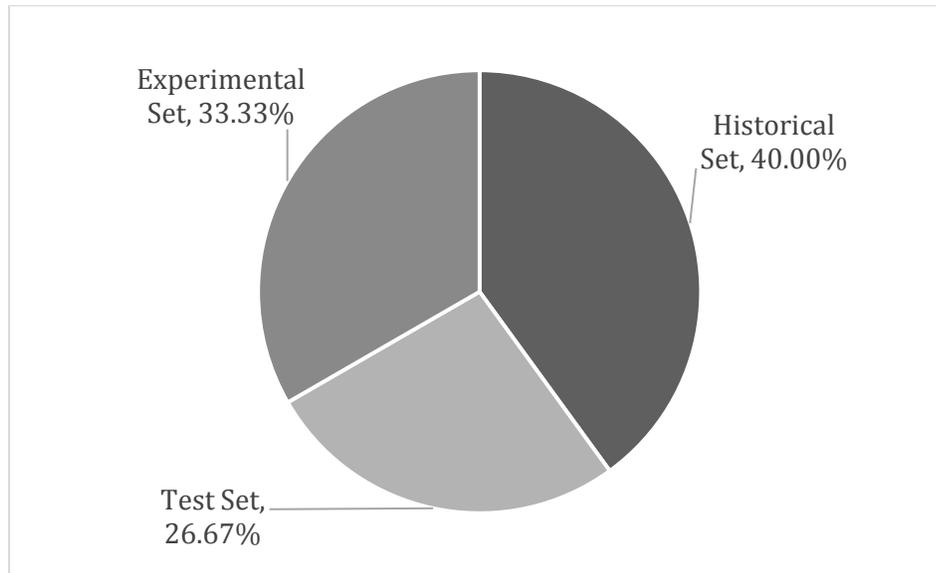

*Figure 2*. Segmentation of the data set.

## Training Phase

A training phase was necessary to assign weights to each attribute so as to minimize error of estimation. Since each experiment was run with multiple values of *k* nearest neighbors, it was necessary to repeat the training phase six times for *k* values from one to six.

As discussed in the "previous work" section, many different approaches have been proposed to find these weights. This study relied primarily on a simple stochastic technique, following Bergstra and Bengio (2012), in which numerous random combinations of weights were tested and the combination with the lowest amount of error retained. Since this solution only represented an approximate minimum of the error function, a numeric optimization algorithm was then applied to "fine tune" the weights and find a true local minimum.

In the stochastic phase, weights were randomly chosen on the interval $[0, 1]$. This range was chosen because it provides a theoretically infinite number of ratios between the weights, while still being a bounded interval, and thus avoiding potential problems with floating point overflow. Also, since Python's built in random number generator returns numbers on this range, several calculations were saved in each iteration.

For each combination of weights, the cost was estimated for every job in the training set, based on the historical set. The mean average percentage error for all of the estimates was then calculated as a fitness criterion. After 22,500 iterations, the combination of parameters with the lowest mean average percentage error (MAPE) was retained and the fine tuning procedure was invoked.

The fine tuning procedure used the *scipy.optimize.fmin* function from the open source Scipy library. This function uses Nelder's & Mead's (1965) downhill simplex algorithm to minimize an unconstrained multivariate function ("scipy.optimize.fmin", 2013).

This algorithm has the advantage of not needing to compute any first or second derivatives of the function, a critical feature given the discontinuous nature of a $k$-NN error function. However, it is not a particularly fast algorithm and it sometimes gets stuck on local minima of complex functions. Thus, in this application it was more suited to refining solutions generated by the stochastic method. The fine tuning procedure was allowed to run for 2,500 iterations ($O(2)$ days of computer time) or until it converged within the limits of hardware precision.

## Trial 1 – The Basic Method

In the first trial, jobs in the test set were estimated in chronological order from oldest to newest over a period of 83 weeks. All jobs in the training set that were at least 30 days older than the job currently being estimated were added to the database and used to prepare the estimate. This simulated a real firm's use of the method, where jobs over a month old would have already been invoiced, giving the company data on their true costs. This trial established baseline performance measures for the method.

Values of $k$ from one to six were tested. Higher $k$ values require increasingly more computer resources during the training phase, since the time requirements go up geometrically as $k$ increases. Also, smaller $k$ values are more applicable to a non-homogenous data set, where there may only be a few historical jobs of a particular class. For these reasons, $k$ values of six or less were considered to be the most suited to a practical implementation of the method.

Additionally, one of the supposed advantages of lazy learning algorithms like $k$-NN is the fact that the accuracy tends to increase as more information is added to the system (Illa, Alonso, & Marré, 2004). If this effect was present, then the trend in average error over time as more jobs were added should be negative. To see whether there is evidence to believe this is occurring, the following hypothesis was tested:

> $H_0$:  **The trend in absolute percent error as a function of time = 0 as jobs are added to the system over time.**
>
> $H_1$: **The trend in absolute percent error as a function of time ≠ 0 as jobs are added to the system over time.**

Trend was approximated by the slope of the least squares regression line through the average error for each week of the trial as a function of time.

## Trial 2 – Without Attribute Weights

The training phase is the most computationally intensive part of this estimating method. Trial 2 was intended to provide insight on whether a training phase to find the optimal attribute weights is worthwhile. Trial 1 was duplicated with the difference that all attributes of the weights were set equal to one (equally important). The hypothesis tested was:

$H_0$: *The absolute percentage error after training is equal to the mean percentage error with no training.*

$H_1$: *The absolute percentage error after training was lower than the mean percentage error with no training.*

# RESULTS

The mean absolute percentage error (MAPE) and third quartile percentage error (Q3APE) for each trial at each of the six $k$ values is summarized in Table 1. While MAPE was used as the primary fitness criterion in this study, Q3APE is also a good measure of the method's effectiveness, since it is more robust than the mean with respect to extreme estimates. Furthermore, in practice, a methodology that produced good results 75% of the time would be considered useful, provided there were some way to prescreen jobs that were likely to fall in the other 25% so freight forwarders could manually prepare the estimates. Further descriptive statistics of the error distributions of each trial are contained in the appendix.

Table 1

*Overall performance for both trials*

| $k$ | Trial 1 (Trained Attribute Weights) | | Trial 2 (No Attribute Weights) | |
|---|---|---|---|---|
| | MAPE | Q3APE | MAPE | Q3APE |
| 1 | 117.82% | 63.53% | 113.92% | 56.25% |
| 2 | 86.38% | 62.58% | 109.56% | 49.99% |
| 3 | 90.49% | 62.58% | 101.71% | 49.98% |
| 4 | 89.13% | 45.90% | 149.35% | 56.75% |
| 5 | 87.60% | 45.51% | 99.82% | 49.99% |
| 6 | 87.75% | 45.83% | 99.56% | 49.99% |

*Note.* MAPE = Mean absolute percentage error. Q3APE = Third quartile absolute percentage error.

a.

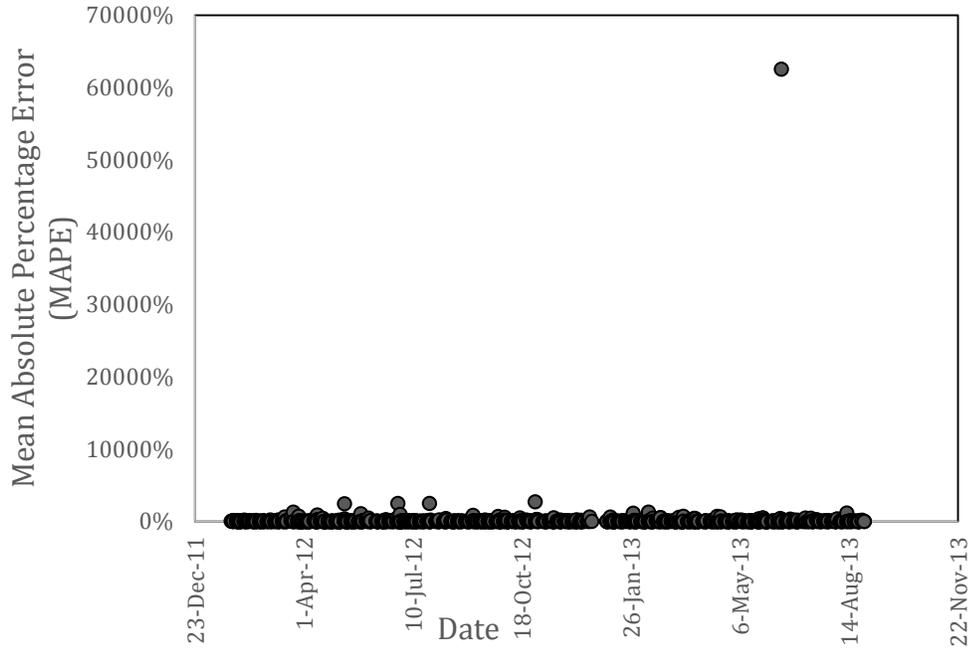

b.

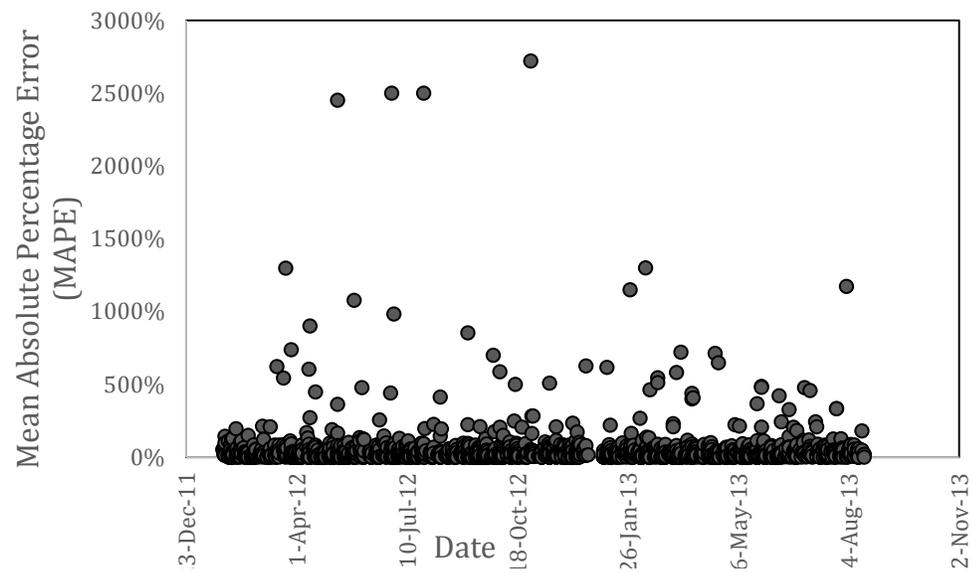

*Figure 3*. Plot of MAPE as a function of time, Trial 1, *k* = 4.  a) One outlier has an MAPE several orders of magnitude worse than any other job.  b) The same data set with the outlier removed.  The plots were similar for other values of *k*.

The overall errors when the method was applied to estimate costs for the entire data set were fairly large. However, it seems likely that the slope was biased by several abnormally high estimates, which can be seen on the scatter plot in Figure 3.

One job in particular had an especially high estimated cost. Upon examining the database, it was discovered that the job, which consisted of two standard pallets sent to Shanghai in June 2013, was one of only three consignments to China in the database. One of the other two consignments was a single envelope, which was most likely sent by air mail. Since this envelope job was one of the nearest neighbors used for the estimate, it caused the two-pallet job to be estimated extremely high. This shows one way that the method can break down when used for consignments to areas that receive few shipments.

The outlying job was excluded from the results for all further analysis.

## Trial 1

The average change in absolute percentage error over time was approximated by fitting a regression line to the absolute percent error versus the date and finding the slope. The slope coefficients for each value of $k$ are summarized in Table 2. Unfortunately, there is no justification at the 95% confidence level to conclude that the error decreased over time.

Table 2

*Hypothesis Test on Slope of Regression Line,  Absolute Percent Error vs. Time*

| $k$ | Slope | Standard Error | t Stat | 95% CI | |
|---|---|---|---|---|---|
| 1 | 2.42E-04 | 0.1137 | 0.1137 | -0.0039 | 0.0044 |
| 2 | -2.28E-04 | 0.0009 | -0.2618 | -0.0019 | 0.0015 |
| 3 | 7.53E-06 | 0.0009 | 0.0084 | -0.0018 | 0.0018 |
| 4 | -7.34E-05 | 0.0002 | -0.3089 | -0.0005 | 0.0004 |
| 5 | -3.55E-05 | 0.0002 | -0.1570 | -0.0005 | 0.0004 |
| 6 | 9.65E-05 | 0.0003 | 0.3835 | -0.0004 | 0.0006 |

*Conclusion*:  Fail to reject null hypothesis at 95% confidence level for all values of $k$.

## Error Analysis by Consignment Type

The following analyses were conducted only on the output with $k = 5$ (the lowest error trial).  The extreme outlier mentioned above was discarded.  As can be seen in Table 3, the MAPE appears to increase as the load size gets bigger.  In particular, consignments smaller than one standard pallet are estimated with significantly more accuracy than larger consignments.

Table 3

*Mean absolute percentage error by load size*

| Load Size | n | MAPE | SE | 95 % CI | |
|---|---|---|---|---|---|
| Less than 1 standard Pallet | 740 | 33.64% | 2.62% | 28.49% | 38.79% |
| At least one standard pallet but less than half load | 601 | 51.06% | 4.36% | 42.49% | 59.62% |
| Half Load and above | 314 | 83.13% | 16.66% | 50.35% | 115.91% |

*Note*.  SE = standard error of the mean.  CI = confidence interval.

The amount of error also seems to vary by the collection and delivery countries (Table 4). For the most part, consignments within Europe had much lower estimation error than consignments which are collected or delivered in the rest of the world. Imports had a slightly higher relative estimation error than exports, but the difference is not statistically significant at the 95% confidence level.

Table 4

*Mean absolute percentage error by collection and delivery region*

| Delivery Region | n | MAPE | SE | 95% CI | |
|---|---|---|---|---|---|
| Ireland | 807 | 55.27% | 5.57% | 44.33% | 66.21% |
| United Kingdom (exc. North Ireland) | 193 | 92.61% | 21.13% | 50.94% | 134.28% |
| Other European Union | 638 | 30.81% | 2.43% | 26.03% | 35.59% |
| Other Europe | 16 | 36.20% | 6.55% | 22.25% | 50.15% |
| Total Europe (inc. Ireland) | 1654 | 50.01% | 3.81% | 42.52% | 57.49% |
| Rest of the World | 21 | 86.45% | 32.53% | 18.60% | 154.31% |
| Total Export | 868 | 46.00% | 5.16% | 35.88% | 56.11% |

| Collection Region | n | MAPE | SE | 95% CI | |
|---|---|---|---|---|---|
| Ireland | 872 | 44.66% | 4.94% | 34.97% | 54.36% |
| United Kingdom (exc. North Ireland) | 327 | 46.12% | 5.91% | 34.50% | 57.75% |
| Other Europe (inc. Other EU) | 463 | 54.83% | 6.10% | 42.85% | 66.81% |
| Total Europe (inc. Ireland) | 1662 | 47.78% | 3.31% | 41.29% | 54.27% |
| Rest of the World | 13 | 392.84% | 233.30% | -115.46% | 901.15% |
| Total Import | 803 | 56.76% | 5.80% | 45.37% | 68.14% |

*Note.* One extreme data point was removed. The total numbers of imports and total exports do not balance because 4 intra-country consignments were included in the data set. SE = standard error of the mean. CI = confidence interval.

The error for exports to the United Kingdom is higher than that for consignments to other regions in Europe. This is likely attributed to the fact that, in the company being studied, routine consignments to the UK are handled by a different business unit, while the freight forwarding department handles unusual consignments and those with special delivery instructions (Straight, 2013). Thus, "unusual" exports to the UK may be over represented in this data set.

Analyzing MAPE by total distance traveled revealed neither patterns nor statistically significant differences at the 95% confidence level.

Because the error varies over different classes of consignments, it might prove worthwhile to partition the data set by consignment type prior to training the algorithm, thus generating different sets of attribute weights for each class of jobs. While time constraints precluded such a trial in the current study, it might be an interesting line of inquiry for future work.

## Trial 2

The bar graph in Figure 4 shows the MAPE at each $k$ value with and without a training phase. The null hypothesis that the mean average percentage error was equal before and after training was tested with a paired t-test at a 95% significance level. The results are summarized in

Table 5. There is no significant evidence to training improves accuracy at $k$ values below 5.

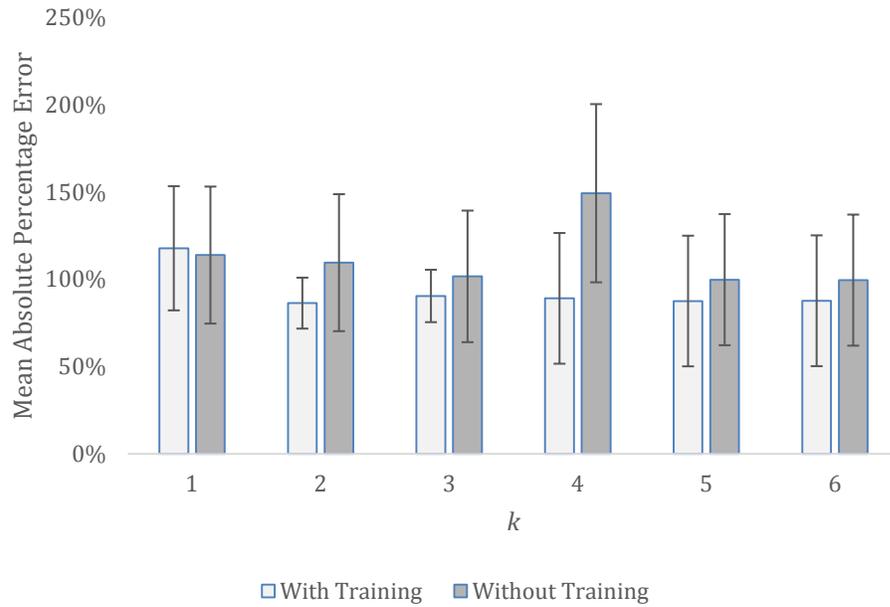

*Figure 4.* Mean absolute percentage error with and without a
training phase.  Error bars represent the standard error of the mean.

Table 5

*Mean absolute percentage error with and without training*

| | MAPE | | | |
| *k* | **With Training** | **Without Training** | **Conclusion** | **Significance** |
|---|---|---|---|---|
| 1 | 117.8% | 113.9% | Fail to reject $H_0$ | .936 |
| 2 | 86.4% | 109.6% | Fail to reject $H_0$ | .537 |
| 3 | 90.5% | 101.7% | Fail to reject $H_0$ | .774 |
| 4 | 89.1% | 149.4% | Fail to reject $H_0$ | .084 |
| 5 | 87.6% | 99.8% | Reject $H_0$ | .003 |
| 6 | 87.8% | 99.6% | Reject $H_0$ | .001 |

*Note.* The null hypothesis was that absolute percentage error is the same
with training as without training.

# DISCUSSION

## Overall Effectiveness of the Method

The results of the first trial imply that the methodology does seem to have value for certain classes of jobs. In particular, it works relatively well on consignments that are smaller than a standard pallet (MAPE= 33.64%) and exports to countries within the European Union (MAPE = 30.81%). The important issue, however, is how this level of accuracy compares to other estimation methods applied to the same data set.

Multiple linear regression is a popular technique for predictive analytics, and is fairly simple to implement with modern computer software. A stepwise linear regression model was built from the data set using SPSS (*IBM SPSS Statistics*, 2013). The variables for date and load size were used, as well as total distance traveled, collection country, and delivery country. Full output from the model is contained in Appendix 2.

The regression model had a lower MAPE than the analogy based model, but a higher Q3APE (Table 6). This implies that, while the linear model is less sensitive to outliers, probably because SPSS is effective at detecting and trimming them, it is not as accurate for estimating the cost of a typical job. A paired T-test failed to reject the null hypothesis that the mean of the differences between the two methods is zero (95% CL, p-value = 0.668). In other words, there is no significant difference in the accuracy of the two methods.

Table 6

*Comparison of different forecasting methods*

| | n | MAPE | 95% CI on MAPE | | Q3APE |
|---|---|---|---|---|---|
| Manual estimation by experienced freight forwarder | 1673 | 13.23% | 12.26% | 14.21% | 15.31% |
| Analogy based method | 1679 | 87.60% | 14.26% | 160.94% | 45.51% |
| Multiple linear regression | 1679 | 71.51% | 63.72% | 79.31% | 61.79% |
| Combination of analogy and regression methods | 1679 | 36.14% | 33.17% | 39.10% | 42.46% |

*Note.* MAPE = mean absolute percentage error. Q3APE = third quartile absolute percentage error. CI = confidence interval. Only the subset of jobs (out of the same 1679) with revenue information could be used for the manual estimate.

Estimation by analogy and multiple linear regression are but two of the many predictive techniques available, and some other might provide more accuracy than either. It is likely that the best possible technique would use a weighted average of multiple methods, as combined forecasts tend to be more accurate; unless their forecast errors are highly correlated (*Combining Forecast Results* in Wilson & Keating, 2009). The correlation coefficient between the errors of the analogy method and the errors of regression method was only 0.0748. Given that most of the outliers are overestimates, a forecast was used that estimated the job as the minimum estimate from each method. This technique improved both MAPE and Q3APE compared to either the analogy based method or the linear regression method by themselves (Table 6). It is possible that combining the forecasts differently or bringing in the results of other methods could bring the error down even further.

An even more important question than how the analogy based method compares with other predictive techniques is how it compares to manual estimation of job costs. Not surprisingly, both the MAPE and the Q3APE of manual estimation are considerably lower than either the analogy based method or the regression model (Table 6). An explanation of how the amount of error was estimated for manual techniques is contained in Appendix 3.

Even though manual estimates are clearly more accurate, there might still be reasons for a firm to implement an analogy based method. Compiling freight estimates can be a labor intensive process. The additional costs imposed by an imperfect estimating system might be offset, in some cases, by savings in labor. That is, if it costs some amount, $C_e$, to prepare each estimate, then if $C_e$ was above a certain level, it would become worthwhile to switch to an automated system. The value of this point of indifference would be of some interest. In general, since $C_e$ is the difference between the expected value of profit per job using manual estimation and the expected value of profit per job using automated estimation:

$$C_e = \overline{P_0} - \overline{P_1} \qquad (4)$$

*where:*

$\overline{P_0} =$ *The average gross profit per job using manual estimation*

$\overline{P_1} =$ *The average gross profit per job using a predictive method*

All of the random variables involved are non normal, which makes a closed form analytical solution difficult. However, a Monte Carlo simulation, using the actual distributions of historical costs and error percentages for the three methods, was used to estimate the value of $C_e$ for both the analogy based method and the combined analogy/regression method (

Table 7). Labor costs specifically for freight forwarders were not available. However, the median labor cost in Ireland for "industry, contracting, and service workers" was €28.29/hour in 2008. Taking this as a typical labor cost, the analogy based method would dominate if it took the company more than 1.5 hours to complete an estimate ("Labour cost, wages and salaries, direct remuneration", 2008).

The estimate of $C_e$ was somewhat sensitive to the target gross margin and the average number of bidders per job, but a realistic range of these values still yielded values of $C_e$ such that the methods, particularly the combined method, might be viable in a practical application (Figure 5).

Table 7

*Labor cost at which firms would be indifferent between estimating methods*

| Method | Estimated $C_e$ | 95% CI | |
|---|---|---|---|
| Analogy based method | €42.45 | €40.74 | €44.16 |
| Combination of analogy based and regression methods | €34.45 | €32.96 | €35.95 |

*Note*: Monte Carlo simulation with 25,000 trials. Assumptions: jobs let in a simple auction with 3 bidders. All bidders strove for a 15.1% gross margin. CI = confidence interval.

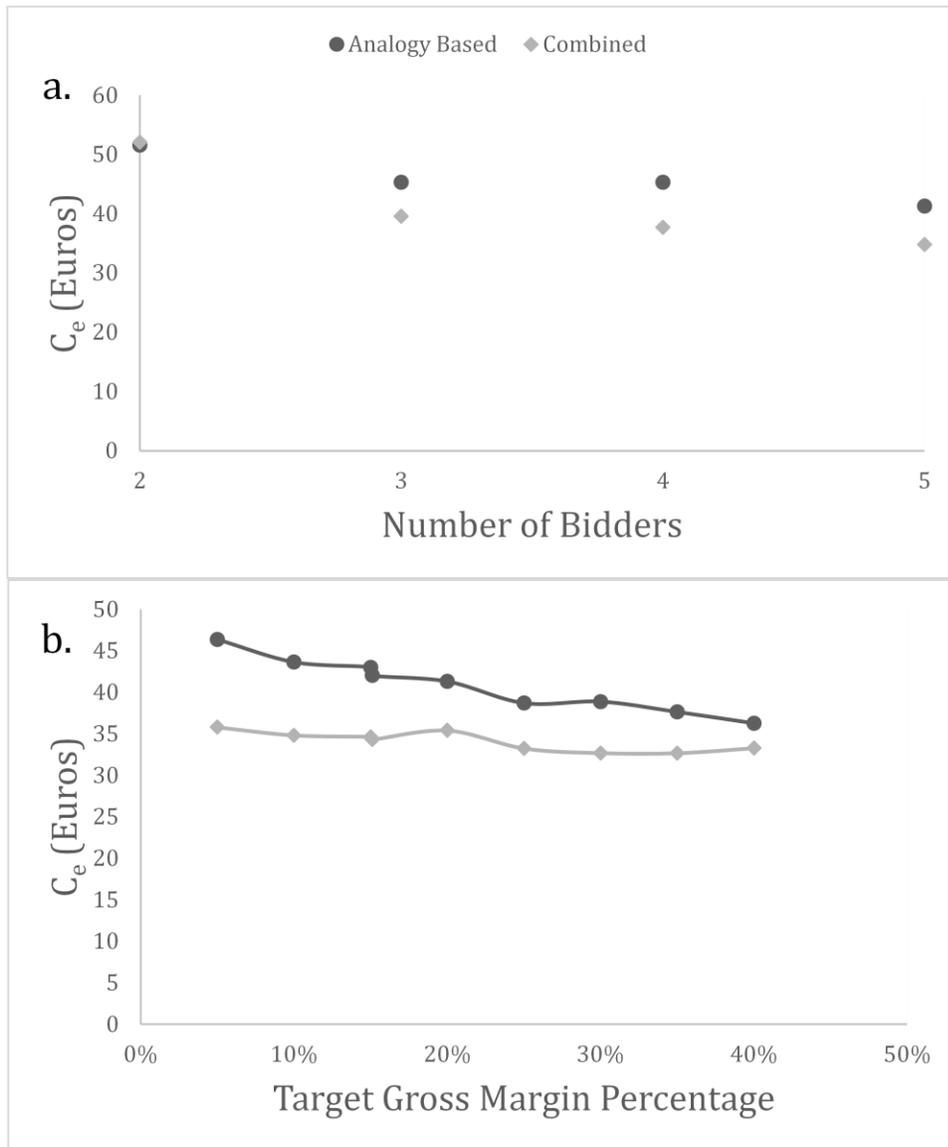

*Figure 5*. Sensitivity analysis of point of indifference in estimate labor cost for number of bidders and target gross margin percentage.

## Lessons for Implementing an Analogy Based Method

A predictive analytic method is only as good as the data set it uses. In this case, the data set had numerous incomplete records and other issues. This is probably one of the major reasons that the overall accuracy suffered. If a firm were to implement the method, the first step would be to design a data entry interface with validation features that would force employees to enter complete information about every job, including full addresses, total weight, and overall dimensions. The importance of capturing complete and accurate data cannot be overstated.

The need for a training phase should be assessed in any new implementation. Training this method required a substantial investment of computer time, for only a small increase in accuracy. In fact, the improvement was only statistically significant at $k$ values over 4. The choice $k$ value itself does affect the accuracy of the method. Even if a full training phase is foregone, different $k$ values should still be tested to find the best one for a given data set.

The final lesson is that, so far at least, no automated method is as accurate as having a trained freight forwarder creating a bottom up manual estimate, calling subcontractors for quotes as necessary. Input from practicing freight forwarders was invaluable to preparation of this paper. However, it is interesting to note that these individuals were not always right. For example, all three of the freight forwarders interviewed in the field stressed the importance of fuel prices as the most important factor in international freight costs and claimed that "our costs are always going up because the price of oil is always going up" (Cook, O., Conway, A., & Vahey, M, personal communication, 2013). However, there is actually a weak negative correlation between oil and freight prices (Figure 6. Normalized oil price and freight cost per truckload over time. Coefficient of correlation = -0.33767.). In fact, the normalized freight costs for this company display a slight downward trend.

This example illustrates the point that even seasoned professionals might display bias in their observations of the international freight market.  In this case, they were probably succumbing to an availability bias, caused by the easy availability of information on rising oil prices in popular culture and the media.  There is no evidence to support the relationship, however.  Anyone seeking to build a predictive model for this market would do well to consult practitioners, but should base the model itself on mathematical and statistical analysis, based on the best available data.

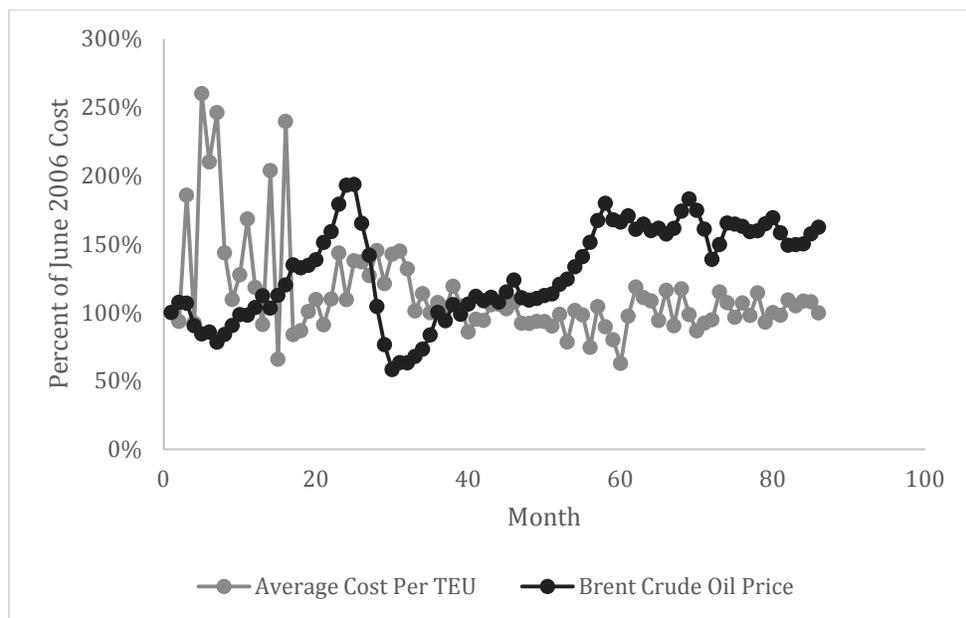

*Figure 6.*  Normalized oil price and freight cost per truckload over time. Coefficient of correlation = -0.33767.  Oil prices are courtesy of the Energy Information Administration (2014).

# SUGGESTIONS FOR FUTURE RESEARCH

Two areas seem particularly interesting for future research: scalability of the method and sampling policy within the estimating method.

## Scalability

The data set used for this study is large compared to most published data sets used to test analogy based estimating methods (Mair, Shepperd, & Jørgensen, 2005). However, it was generated by a relatively small volume freight forwarder.  It would be interesting to explore how the system would scale up for use by a larger operation.  The training phase, in particular, is already rather time consuming at the present scale.  The process is already parallelized, and there is no immediate limitation on the number of processing nodes that could be added.  However, in the best case, adding computers will only give a linear increase in processing power.  Processing requirements for the estimator scale linearly as jobs are added to the database because each new job adds one more attribute distance which needs to be calculated.  However, processing requirements for the training phase go up quadratically as new jobs are added, leading to diminishing returns of adding more hardware.  One answer would be to partition the data based on one or more factors and train the method for each class of job separately.  Besides adding scalability, this might possibly add accuracy for estimates of jobs within each class.

**Sampling Policy**

This study has focused on ways to improve the basic $k$-NN estimating methodology and on ways to identify estimates that are likely to be in error. A more sophisticated methodology would avoid the root causes of error while the estimate was being made. The following features seem promising, but there was insufficient time to implement and test them in the current study.

Two of the more pernicious causes of error are overrepresentation and underrepresentation of job data. Figure 7[3] shows the ideal case for $k$-NN estimation. A prospective job, $E$, falls into a cluster which contains at least $k$ similar jobs, spaced equally around it in $n$-space. The estimate is essentially an interpolation between their costs.

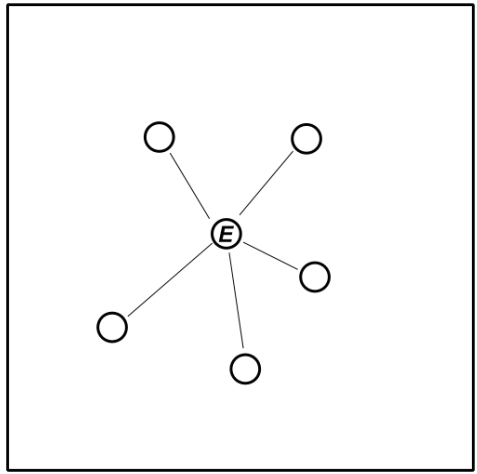

*Figure 7.* The job to be estimated roughly falls within a cluster and is roughly equidistant from $k$ nearby jobs in $n$-space. This figure is drawn in two dimensions for simplicity.

[3] These figures were inspired by those in Mitchell (1997).

Overrepresentation occurs when a particular consignment profile is overrepresented in the database. For instance, a particular consignee might receive several nearly identical shipments over a period of time at around the same cost. Whenever a prospective job to the area was estimated, these redundant jobs would probably be overrepresented in the $k$-NN, and potentially bias the estimate. This case is illustrated in Figure 8. A possible solution would be an under-sampling policy that ignores some of the redundant jobs, instead using other nearby jobs as nearest neighbors.

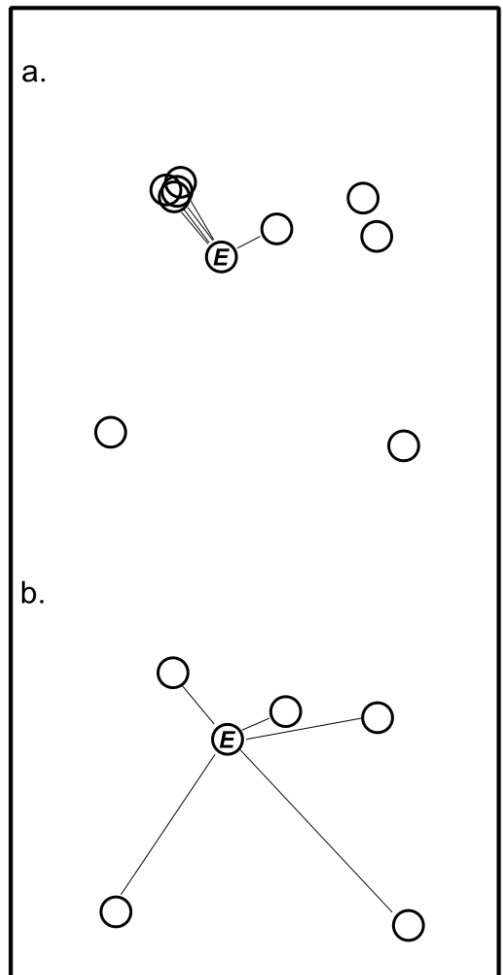

*Figure 8.* a) The job to be estimated falls near several previous jobs in n-space which are nearly identical and bias the estimate.  b) The solution is to under sample the redundant jobs.

Underrepresentation is exemplified by the job to Shanghai, discussed in the results section, which biased the results of Trial 1.  There were only two other jobs in the database with delivery addresses in the area, and one was significantly different in size, causing the estimate to be in error by a large amount.  A similar situation is shown in Figure 9.  A possible solution is to use only the jobs that are within the cluster as members of the nearest neighbors, then to create synthetic data to fill the remaining spots.

This is conceptually similar to how the SMOTE algorithm works (Chawla, Bowyer, Hall, & Kegelmeyer, 2011). Techniques for creating synthetic data for *k*-NN applications have been described by various authors, including Li, Xie, and Goh (2009).

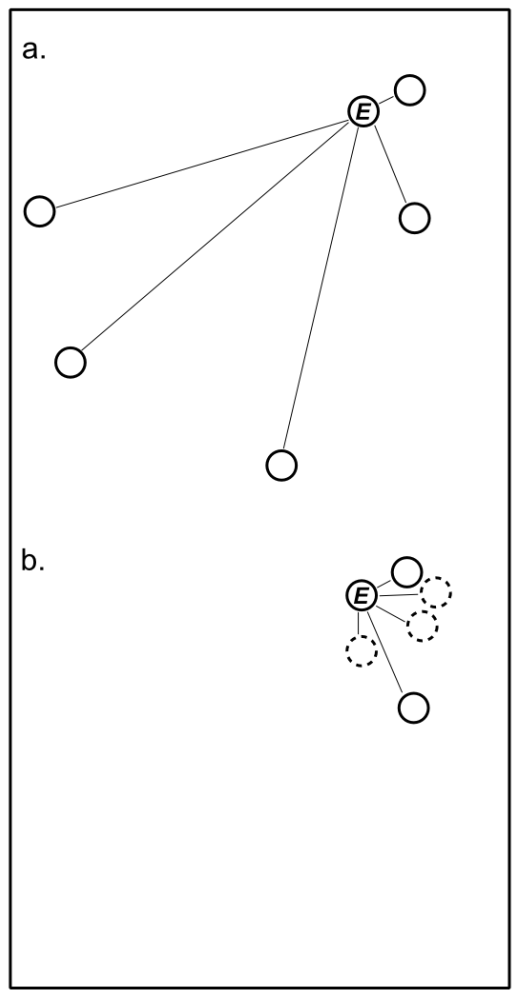

*Figure 9.* a) There are less than *k* nearby jobs, causing some less similar jobs to be included in the *k* nearest neighbors. b) The solution is to oversample the nearby jobs by creating synthetic records.

A related sampling problem is illustrated in Figure . It illustrates the case where a job falls outside any cluster. Since *k*-NN is not capable of extrapolation, any estimate made would be unreliable. The method should have the ability to detect this case, and return an error code instead of an estimate. In *n*-space, this could most accurately be done by measuring the angle between the *k* nearest neighbors, as angles are more stable than distances in high dimensional spaces (Kriegel, Kröger, & Zimek, 2009). If the total angle was too acute, the estimating procedure would fail gracefully and the operator would know to manually prepare an estimate.

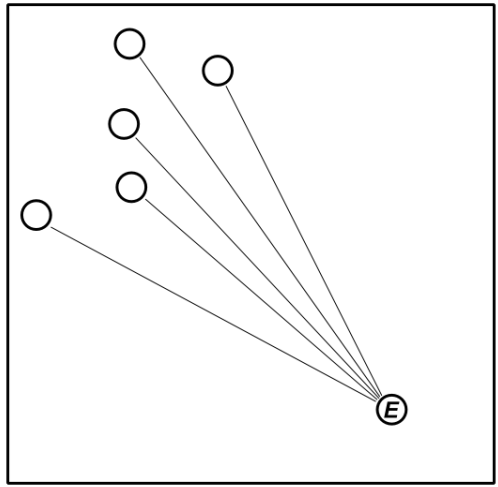

*Figure 10.* The job to be estimated falls far from any previous jobs. An accurate estimate cannot be made and the method should fail gracefully.

# APPENDIX 1 – DESCRIPTIVE STATISTICS FROM THE TRIALS

**Trial 1 (Base Case)**

|  |  | Error (€) | Error (%) | Abs. Error (€) | Abs. Error (%) |
|---|---|---|---|---|---|
| $k = 1$ | Mean | 445.88 | 87.9% | 569.56 | 117.8% |
|  | Std. Dev. | 8,518.07 | 1460.1% | 8,510.69 | 1458.0% |
|  | Min | (2,113.21) | -97.8% | - | 0.0% |
|  | Q1 | (67.75) | -23.4% | 20.00 | 9.7% |
|  | Median | - | 0.0% | 82.81 | 29.9% |
|  | Q3 | 110.00 | 40.9% | 237.00 | 63.5% |
|  | Max | 343,824.92 | 56220.0% | 343,824.92 | 56220.0% |
| $k = 2$ | Mean | 261.38 | 56.0% | 390.51 | 86.4% |
|  | Std. Dev. | 1,956.86 | 601.5% | 1,935.22 | 597.9% |
|  | Min | (2,084.10) | -99.6% | - | 0.0% |
|  | Q1 | (66.51) | -24.8% | 17.50 | 8.4% |
|  | Median | - | 0.0% | 76.58 | 29.4% |
|  | Q3 | 96.02 | 34.4% | 235.48 | 62.6% |
|  | Max | 57,149.13 | 23137.3% | 57,149.13 | 23137.3% |
| $k = 3$ | Mean | 280.20 | 60.6% | 406.74 | 90.5% |
|  | Std. Dev. | 2,013.86 | 618.8% | 1,992.14 | 615.2% |
|  | Min | (2,084.10) | -99.6% | - | 0.0% |
|  | Q1 | (64.79) | -23.9% | 16.34 | 8.4% |
|  | Median | - | 0.0% | 75.00 | 29.2% |
|  | Q3 | 96.02 | 35.5% | 235.35 | 62.6% |
|  | Max | 57,149.13 | 23137.3% | 57,149.13 | 23137.3% |
| $k = 4$ | Mean | 214.61 | 63.9% | 321.50 | 89.1% |
|  | Std. Dev. | 3,022.53 | 1535.3% | 3,013.03 | 1534.0% |
|  | Min | (1,519.80) | -99.8% | - | 0.0% |
|  | Q1 | (50.00) | -19.5% | 8.00 | 3.3% |
|  | Median | - | 0.0% | 53.52 | 20.3% |
|  | Q3 | 56.55 | 21.2% | 166.97 | 45.9% |
|  | Max | 114,609.04 | 62552.7% | 114,609.04 | 62552.7% |

*Trial 1 (Base Case, cont.)*

|  | | Error (€) | Error (%) | Abs. Error (€) | Abs. Error (%) |
|---|---|---|---|---|---|
| *k* = 5 | Mean | 202.98 | 62.4% | 317.41 | 87.6% |
| | Std. Dev. | 3,015.46 | 1534.4% | 3,005.57 | 1533.2% |
| | Min | (6,663.72) | -99.8% | - | 0.0% |
| | Q1 | (50.51) | -19.0% | 7.89 | 3.6% |
| | Median | - | 0.0% | 54.00 | 20.1% |
| | Q3 | 57.26 | 21.5% | 162.06 | 45.5% |
| | Max | 114,609.04 | 62552.7% | 114,609.04 | 62552.7% |
| *k* = 6 | Mean | 190.17 | 62.7% | 300.14 | 87.8% |
| | Std. Dev. | 2,952.70 | 1536.3% | 2,943.55 | 1535.1% |
| | Min | (2,084.10) | -99.8% | - | 0.0% |
| | Q1 | (50.01) | -18.5% | 6.83 | 3.3% |
| | Median | - | 0.0% | 53.34 | 19.9% |
| | Q3 | 57.30 | 20.9% | 160.36 | 45.8% |
| | Max | 114,609.04 | 62552.7% | 114,609.04 | 62552.7% |

***Trial 2 (Without Training)***

|  |  | Error (€) | Error (%) | Abs. Error (€) | Abs. Error (%) |
|---|---|---|---|---|---|
| *k* = 1 | Mean | 346.7918619 | 85% | 481.2461307 | 114% |
|  | Std. Dev. | 3548.678645 | 1609% | 3532.947853 | 1608% |
|  | Min | -5441.913677 | -98% | 0 | 0% |
|  | Q1 | -63.5 | -23% | 16 | 7% |
|  | Median | 0 | 0% | 70.00879736 | 26% |
|  | Q3 | 75.05070541 | 31% | 208.7057573 | 56% |
|  | Max | 114609.0436 | 62553% | 114609.0436 | 62553% |
| *k* = 2 | Mean | 258.7344336 | 83% | 376.1099026 | 110% |
|  | Std. Dev. | 3240.845639 | 1609% | 3229.322049 | 1608% |
|  | Min | -2084.1 | -100% | 0 | 0% |
|  | Q1 | -54.65555283 | -21% | 10 | 5% |
|  | Median | 0 | 0% | 62 | 22% |
|  | Q3 | 68.643448 | 23% | 185.2520589 | 50% |
|  | Max | 114609.0436 | 62553% | 114609.0436 | 62553% |
| *k* = 3 | Mean | 239.1911438 | 75% | 356.6579544 | 102% |
|  | Std. Dev. | 3047.361892 | 1546% | 3035.849189 | 1544% |
|  | Min | -2084.1 | -100% | 0 | 0% |
|  | Q1 | -54.38280234 | -21% | 10 | 5% |
|  | Median | 0 | 0% | 62.47749944 | 22% |
|  | Q3 | 68.643448 | 24% | 185.3430589 | 50% |
|  | Max | 114609.0436 | 62553% | 114609.0436 | 62553% |
| *k* = 4 | Mean | 477.4434275 | 121% | 604.8509198 | 149% |
|  | Std. Dev. | 7581.468858 | 2093% | 7572.363928 | 2091% |
|  | Min | -5808.329433 | -98% | 0 | 0% |
|  | Q1 | -56.59727052 | -22% | 17.19525536 | 8% |
|  | Median | 0 | 0% | 70 | 26% |
|  | Q3 | 85.74916527 | 33% | 208.7401249 | 57% |
|  | Max | 279166.868 | 62553% | 279166.868 | 62553% |

*Trial 2 (Without Training, cont.)*

|  |  | Error (€) | Error (%) | Abs. Error (€) | Abs. Error (%) |
|---|---|---|---|---|---|
| *k* = 5 | Mean | 233.24989 | 73% | 352.3544935 | 100% |
|  | Std. Dev. | 3030.918509 | 1541% | 3019.38359 | 1539% |
|  | Min | -2084.1 | -100% | 0 | 0% |
|  | Q1 | -54.81390957 | -21% | 10 | 5% |
|  | Median | 0 | 0% | 62.53 | 22% |
|  | Q3 | 68.90344603 | 23% | 185.8697223 | 50% |
|  | Max | 114609.0436 | 62553% | 114609.0436 | 62553% |
| *k* = 6 | Mean | 233.5027294 | 73% | 351.1982977 | 100% |
|  | Std. Dev. | 3029.446958 | 1540% | 3018.060824 | 1539% |
|  | Min | -2084.1 | -100% | 0 | 0% |
|  | Q1 | -54.38280234 | -21% | 10 | 5% |
|  | Median | 0 | 0% | 62.47749944 | 22% |
|  | Q3 | 68.643448 | 23% | 185.5268823 | 50% |
|  | Max | 114609.0436 | 62553% | 114609.0436 | 62553% |

**APPENDIX 2 – OUTPUT FROM SPSS MULTIPLE REGRESSION MODEL**

## Model Summary

| | |
|---|---|
| **Target** | Cost |
| **Automatic Data Preparation** | On |
| **Model Selection Method** | Forward Stepwise |
| **Information Criterion** | 58,729.749 |

The information criterion is used to compare to models. Models with smaller information criterion values fit better.

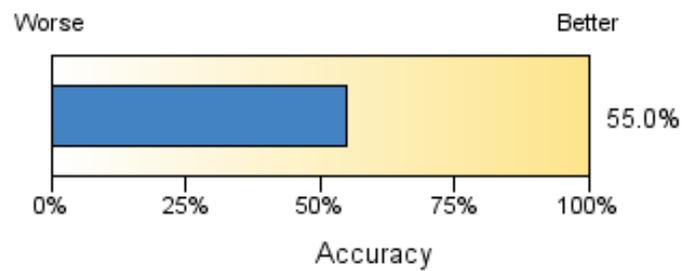

## Automatic Data Preparation

**Target: Cost**

| Field | Role | Actions Taken |
|---|---|---|
| **(CollectionCountry_transformed)** | Predictor | Merge categories to maximize association with target |
| **(CrowDist_transformed)** | Predictor | Trim outliers |
| **(Date_transformed)** | Predictor | Trim outliers |
| **(DeliveryCountry_transformed)** | Predictor | Merge categories to maximize association with target |
| **(LoadSize_transformed)** | Predictor | Trim outliers |

If the original field name is X, then the transformed field is displayed as (X_transformed). The original field is excluded from the analysis and the transformed field is included instead.
One or more records were excluded because of a predictor or target that is missing, a frequency weight that is missing or less than one after rounding, or a regression weight that is missing, negative, or zero.

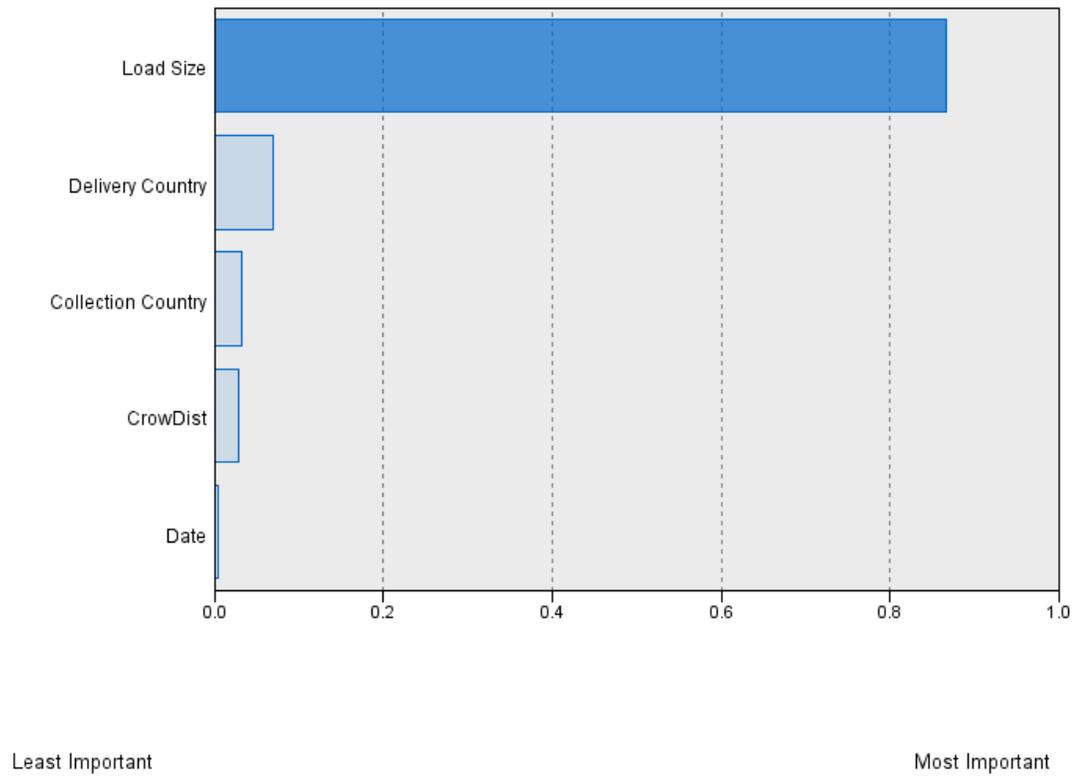

**Predictor Importance**

Target: Cost

Least Important

Most Important

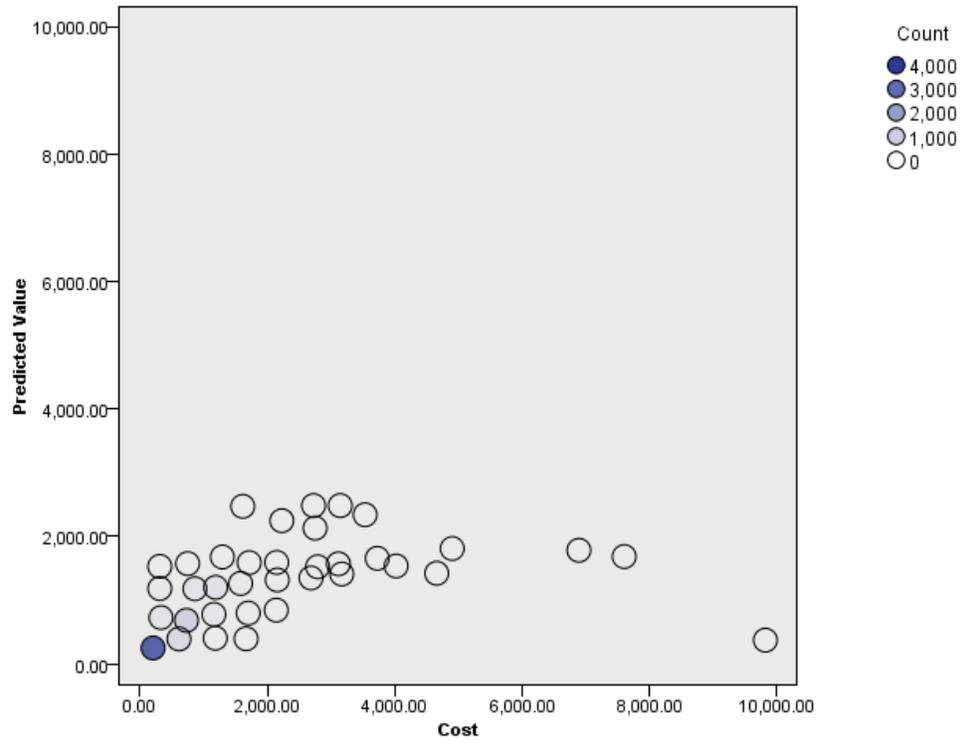

**Predicted by Observed**

Target: Cost

## Residuals

**Target: Cost**

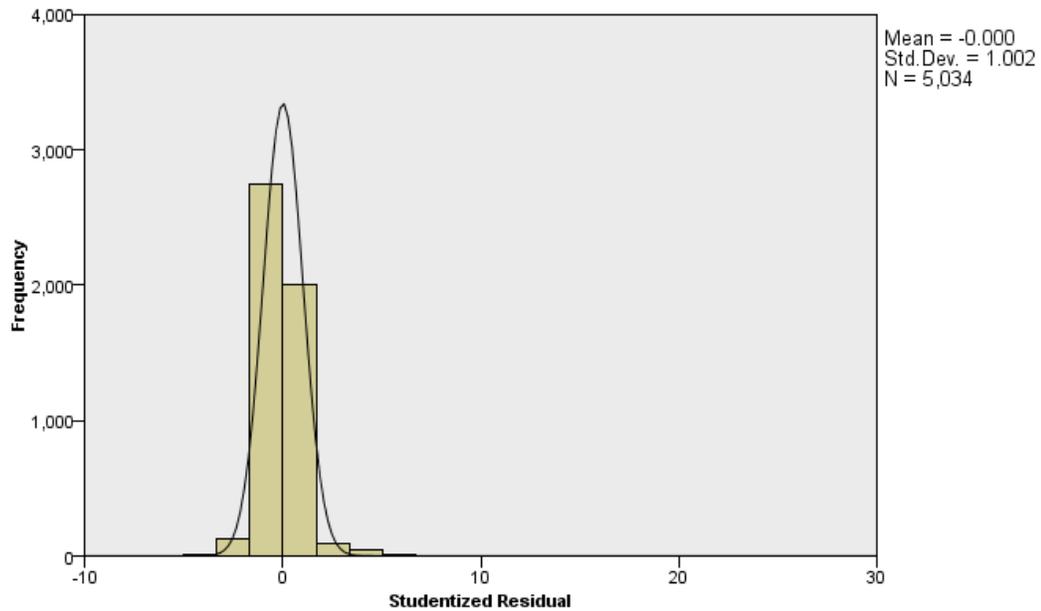

The histogram of Studentized residuals compares the distribution of the residuals to a normal distribution.The smooth line represents the normal distribution.The closer the frequencies of the residuals are to this line, the closer the distribution of the residuals is to the normal distribution.

## Effects

## Target: Cost

| Source | Sum of Squares | df | Mean Square | F | Sig. | Importance |
|---|---|---|---|---|---|---|
| Corrected Model ▼ | 715,437,177.558 | 14 | 51,102,655.540 | 439.538 | .000 | |
| LoadSize_transformed | 503,574,081.622 | 1 | 503,574,081.622 | 4,331.282 | .000 | 0.867 |
| DeliveryCountry_transformed | 40,586,211.239 | 6 | 6,764,368.540 | 58.181 | .000 | 0.070 |
| CollectionCountry_transformed | 18,293,523.327 | 5 | 3,658,704.665 | 31.469 | .000 | 0.031 |
| CrowDist_transformed | 16,507,211.740 | 1 | 16,507,211.740 | 141.980 | .000 | 0.028 |
| Date_transformed | 2,086,959.793 | 1 | 2,086,959.793 | 17.950 | .000 | 0.004 |
| Residual | 583,531,180.378 | 5,019 | 116,264.431 | | | |
| Corrected Total | 1,298,968,357.936 | 5,033 | | | | |

**Model Building Summary**

**Target: Cost**

| | | Step | | | |
|---|---|---|---|---|---|
| | | **1** | **2** | **3** | **4** | **5** |
| **Information Criterion** | | 59,906.246 | 59,263.398 | 58,877.200 | 58,745.708 | 58,729.749 |
| **Effect** | **LoadSize_transformed** | ✓ | ✓ | ✓ | ✓ | ✓ |
| | **DeliveryCountry_transformed** | | ✓ | ✓ | ✓ | ✓ |
| | **CollectionCountry_transformed** | | | ✓ | ✓ | ✓ |
| | **CrowDist_transformed** | | | | ✓ | ✓ |
| | **Date_transformed** | | | | | ✓ |

The model building method is Forward Stepwise using the Information Criterion.
A checkmark means the effect is in the model at this step.

## APPENDIX 3 – ESTIMATION OF ERROR FOR MANUAL ESTIMATES

It was desirable to estimate the general degree of error made when professional freight forwarders originally estimated the costs of jobs in the data set, so as to have baseline with which to compare the analogy based method.  This task was complicated be the fact that, while most records contained information on the actual revenue and costs of the consignments, they did not capture the estimated cost.  Thus, a certain amount of logic was required to derive an estimate of the original error in each of these historic jobs.

First, assume a set of $n$ jobs, and let the following characteristics represent information about these jobs:

$C_i = the\ actual\ cost\ of\ job\ i$

$R_i = the\ actual\ revenue\ from\ job\ i$

$e_i = the\ error\ in\ the\ freight\ forwarder's\ estimate\ for\ job\ i$

Next, assume that the freight forwarder tries to achieve an average target gross margin percentage, $t$.  Assuming that the freight forwarder marks up his estimated cost by this percentage:

$$R_i = (C_i + e_i) \cdot (1 + t)$$

Or, solving for $e_i$:

$$e_i = \frac{R_i - C_i(1+t)}{1+t}$$

Furthermore, we can reasonably assume that, over the long run, the mean error approaches 0, on the grounds that, in an efficient market, a freight forwarder whose estimates were consistently low or high would have trouble keeping his job:

$$\frac{1}{n} \sum_{i=1}^{n} e_i = 0$$

Given these two relations, it was possible to load the data set into a spreadsheet and solve for the value of $t$ which made the average error equal to zero. For the period under consideration, $t$ was approximately 15.1%. Given the values of $e_i$ at this value, the MAPE, Q3APE, and standard deviation of the percentage error could then easily be calculated (Table 8).

Table 8

*Estimated performance of human estimators*

| | |
|---|---|
| Average gross margin percentage (t) | 15.10% |
| | |
| Mean absolute percentage error (MAPE) | 13.02% |
| Third quartile absolute percentage error (Q3APE) | 15.40% |
| | |
| Mean percentage error | 0.00% |
| Standard deviation of percentage error | 24.86% |

# APPENDIX 4 – ATTRIBUTE WEIGHTS AS DETERMINED DURING TRAINING PHASE

Table 9

*Attribute weights as determined during training phase*

| k | Time | Load Size | Collection Location | Delivery Location |
|---|------|-----------|--------------------|--------------------|
| 1 | 0.33034 | 0.71376 | 0.54505 | 0.00373 |
| 2 | 0.84167 | 0.85421 | 0.03379 | 0.00892 |
| 3 | 0.90824 | 0.27774 | 0.03940 | 0.01523 |
| 4 | 0.02588 | 0.18883 | 0.92940 | 0.35002 |
| 5 | 0.01996 | 0.45383 | 0.94084 | 0.48291 |
| 6 | 0.00187 | 0.51252 | 0.01733 | 0.72911 |